\documentclass[sigconf,nonacm]{acmart}
\usepackage{enumitem}
\begin{document}

\title{Where Does the Human End? Creative Agency with Generative AI across Five Years of Chinese Digital Painting}

\author{Yibo Meng}
\authornote{Yibo Meng and Ruiqi Chen contributed equally to this work.}
\orcid{0009-0009-0370-5456}
\affiliation{  \institution{Cornell University}
  \city{NEW YORK}
  \country{USA}
}
\email{yim4007@med.cornell.edu}

\author{Ruiqi Chen}
\authornotemark[1]
\orcid{0000-0001-7255-5353}
\affiliation{  \department{Human Centered Design \& Engineering}
  \institution{University of Washington}
  \city{Seattle}
  \state{WA}
  \country{USA}
}
\email{ruiqich@uw.edu}

\author{Shuheng Cao}
\orcid{0009-0002-1998-396X}
\affiliation{  \institution{University of California San Diego}
  \city{San Diego}
  \state{CA}
  \country{USA}
}
\email{s8cao@ucsd.edu}

\author{Weijia Zhang}
\orcid{0009-0005-2342-6797}
\affiliation{  \institution{University of illinois at Urbana-Champaign}
  \city{Urnana}
  \country{USA}
}
\email{weijia4@illinois.edu}

\author{Chengxi Zang}
\orcid{0000-0002-8244-9551}
\affiliation{  \institution{Cornell University}
  \city{NEW YORK}
  \state{NY}
  \country{USA}
}
\email{chz4001@med.cornell.edu}

\renewcommand{\shortauthors}{Meng et al.}

\begin{abstract}
  As generative AI enters creative work, practitioners must decide where AI assistance ends and human authorship begins. Human-agent interaction (HAI) research has examined AI as a tool, collaborator, consultant, and competitor. The longitudinal problem is how these roles are revised as systems become more capable, public, and economically embedded. We report a five-year interview study with 17 Chinese digital painters, based on annual semi-structured interviews from 2021 to 2025. Participants described recurring but non-uniform patterns of protective resistance, pragmatic task delegation, and, for some, reflective agency repartitioning. Early resistance protected observation, originality, signature, and ownership from AI. Later delegation placed AI in bounded tasks such as references, backgrounds, rough sketches, and client-facing drafts. By 2025, some participants built hybrid workflows around human-only zones, while others described fatigue, precarity, or difficulty locating a remaining human role. Peer norms, emotional climates, and production pressures shaped which delegations felt useful, acceptable, or exhausting. Copyright, authorship, and creative labor remained recurring limits on what participants were willing to delegate. We frame these accounts as longitudinal agency partitioning, the situated work of deciding which stages, responsibilities, values, and claims remain human in creative human-agent interaction. We discuss design implications for revisable agency-boundary controls, provenance scaffolds, and community-facing authorship norms.

\end{abstract}



\maketitle
\pagestyle{plain}

\section{Introduction}
Generative AI is changing parts of creative practice~\cite{shelby2024generative, page2025creative, chung2022artistic}. Across illustration, concept design, music, and literature, AI systems are increasingly described as tools, collaborators, consultants, and competitors~\cite{cetinic2022understanding, caramiaux2022explorers, xu2024application}. This shift matters for human-agent interaction (HAI) because creators must repeatedly decide what competence, originality, autonomy, and responsibility to attribute to agent-like systems whose operations remain opaque~\cite{glikson2020human}. Generative systems can accelerate production and expand ideation. They can also disturb established boundaries of authorship, originality, creative labor, and professional identity~\cite{kyi2025governance, tang2024exploring, knight2024impact, meng2025tracing}. The central question is how artists partition creative agency with AI over time.

Existing research has examined artists' attitudes toward AI~\cite{sikorski2025attitudes, johnston2024understanding, inie2023designing, park2024we}, modes of human-agent collaboration~\cite{rezwana2025human, he2023exploring, schecter2025role, xie2025embodied}, and risks around authorship, credit, and labor~\cite{kyi2025governance, knight2024impact}. These studies show that AI can be valued as a source of inspiration while also being treated as a threat to identity and livelihood. Much of this evidence comes from cross-sectional surveys, short-term studies, or prototype evaluations~\cite{lu2025designing, ma2024drawing, lawton2023tool}. The unresolved HAI issue is how creators revisit earlier judgments as systems become publicly visible, socially normalized, and economically embedded. We address this gap by asking which parts of creative agency artists delegate, which boundaries they preserve, and which values they renegotiate across multiple years.

\textit{Agency partitioning} is the allocation of tasks, responsibilities, values, and claims between human and AI at a given point in creative practice. A painter may reject AI as an author while using it to generate references, backgrounds, or rough client drafts. Another may accept AI for color or layout while reserving the face, sketch, final pass, or signature style for human work. \textit{Agency repartitioning} is a later revision of an earlier allocation. \textit{Longitudinal agency partitioning} is the repeated process through which creators maintain, reopen, or revise these allocations across time. These decisions are practical, social, and moral. They assign responsibility for action, value, and credit. This framing connects tool use with authorship, labor, and peer legitimacy while identifying the specific boundary being negotiated.

We focus on \textit{digital painters}, artists who primarily create two-dimensional visual works through digital media such as illustration, concept design, and game or animation art. The Chinese digital-painting context is analytically important because speed-oriented pipelines in game art, online-literature illustration, comics, animation, and platform-mediated peer circulation made efficiency, visibility, and authorship pressures especially salient. We conducted a five-year qualitative study with 17 Chinese digital painters, based on annual in-depth semi-structured interviews from 2021 to 2025. The 2021 baseline captured awareness of pre-diffusion AI-enabled image tools and early image-generation discourse. Later waves followed the wider public circulation and workplace relevance of prompt-controllable image systems. We ask three research questions. \textbf{RQ1.} How did digital painters describe continuity and change in their relationships with generative AI as prompt-controllable image systems became more publicly visible and professionally relevant after 2022? \textbf{RQ2.} Which agency boundaries recurred across the five-year arc, and which oscillated or transformed? \textbf{RQ3.} How did peer communities and collective sensemaking shape what participants considered delegable to AI?

Our analysis identifies recurring but non-uniform patterns of protective resistance, pragmatic task delegation, and, for some participants, reflective agency repartitioning. Across these patterns, peer norms and emotional climates shaped whether AI use felt useful, acceptable, or exhausting. Copyright, authorship, and creative labor appeared as recurring limits on what participants were willing to delegate to AI.

This study makes three contributions to HAI and HCI. \textbf{Empirically,} our five-year longitudinal interview corpus traces creators' lived experience with generative AI across early awareness, wider circulation, and workplace relevance. \textbf{Conceptually,} we develop longitudinal agency partitioning as the situated, temporal work of maintaining, reopening, and revising agency boundaries. \textbf{For design,} we outline how creative AI systems could support revisable agency boundaries, provenance and authorship claims, and community-level norm negotiation. Grounded in one cohort and context, it shows how creative agency is allocated, revisited, and contested over time.

\section{Related Work}
Prior work connects creative AI through artist adoption and resistance, attributed system roles, and professional identity, linking interaction choices to authorship, labor, and professional self-understanding.

\subsection{Creative Practice, Resistance, and Ethical Boundaries}

Human-computer interaction (HCI) and creativity research examines how artists perceive, adopt, and contest generative AI~\cite{bird2024artists, inie2023designing, tang2024exploring, lovato2024foregrounding}. Creative AI can support faster prototyping~\cite{xu2024application}, broader ideation, and stylistic experimentation~\cite{he2023exploring, chung2022artistic, lu2025designing, chung2021intersection}. Creators often evaluate systems more favorably when AI is framed as inspiration or assistance~\cite{hertzmann2020computers, koch2020art}.

The same literature documents ambivalence and sustained resistance~\cite{cetinic2022understanding, ch2019art, yang2020re, park2024we}. Kawakami and Venkatagiri~\cite{kawakami2024impact} report anxiety about the devaluation of artistic labor. Sikorski et al.~\cite{sikorski2025attitudes} find particularly negative sentiments among students and specialized artists. Bird~\cite{bird2024artists} shows artists resisting narratives that treat AI as a creative equal while facing pressure to use it. Artists respond through refusal, bounded assistive use, and negotiation over consent and credit~\cite{inie2023designing, kyi2025governance}.

Research also identifies structural risks in creative AI practice~\cite{bran2023emerging, brand2021design}. Copyright and data provenance are central concerns~\cite{daniele2019ai+, amato2019ai, zeilinger2021tactical}. Artists object to the use of their work for training without consent and dispute the originality and legitimacy of resulting outputs~\cite{kawakami2024impact, kudless2023hierarchies, lovato2024foregrounding}. Kyi et al.~\cite{kyi2025governance} organize these disputes around consent, credit, and compensation. Recognition therefore becomes a consequential negotiation over reputation, authorship, and market legitimacy.

Aesthetic and labor concerns further complicate creative AI use. Artists describe AI imagery as polished but emotionally hollow~\cite{page2025creative}. They also question whether superficial stylistic reproduction can convey intentionality~\cite{porquet2025copying}. Researchers document precarious labor conditions around technology-mediated creative work~\cite{ming2024labor, knight2024impact, munoz2023identity, mako2022emerging}. Lu~\cite{lu2025research} reports that efficiency gains can pressure compensation rates, especially for freelancers and early-career artists. Bird~\cite{bird2024artists} finds that market pressure can lead artists to adopt AI despite personal opposition.

These studies capture opportunities and risks at specific moments. Our longitudinal design traces how artists revisit these judgments as systems become more capable, public, and embedded in work, while authorship, labor, and copyright remain boundaries around new practices.

\subsection{Agency Attribution in Creative Human-Agent Collaboration}

Prior research examines how artists' experiences vary with the role assigned to AI, including advisor, collaborator, and driver~\cite{canet2022dream, johnston2024understanding, shelby2024generative, darabipourshiraz2025ai}. These roles form a spectrum of \textit{attributed agency}, from minimal attribution in advisory use to stronger attribution when AI acts as a driver~\cite{glikson2020human}. As an advisor, AI supports human decision-making through references, compositional alternatives, or technical enhancements while preserving human authorship~\cite{stork2024computer, bennett2024painting, rezwana2025human, davis2025co}. Hu et al.~\cite{hu2025designing} identify consultant roles in recurring HCI interaction patterns. Schecter and Richardson~\cite{schecter2025role} show that advisory framings can support acceptance by centering human control.

As a collaborator, AI is described as a co-creative partner that stimulates exploration~\cite{lawton2023tool, xie2025embodied, caramiaux2022explorers}. Chung~\cite{chung2022artistic} describes iterative refinement systems that users experience as discovery oriented. Page et al.~\cite{page2025creative} show artists using errors and unexpected outputs as creative stimuli. Driver-like roles create sharper tension around intent and authorship. Artists may recognize local strengths in color, texture, and shading while rejecting AI's claim to holistic style, intent, or authorship~\cite{porquet2025copying, he2023exploring, didion2024did}.

Most work examines short-term encounters or prototypes. Role attribution positions a system within an interaction. Agency partitioning identifies permitted tasks, responsibilities, values, and claims. We examine how these permissions change with social expectations and production pressures across years.

\subsection{Identity Work in Technology Adoption}

Identity work is the discursive and practical effort through which professionals construct, maintain, and revise their occupational selves~\cite{snow1987identity, watson2008managing}. Research in information systems (IS) and HCI shows that technology can trigger identity work by redistributing competence, expertise, and authenticity~\cite{stein2013towards}. Recent creative AI research links these pressures to everyday workflow decisions. Munoz et al.~\cite{munoz2023identity} describe creators re-authoring professional selves through micro-decisions about what to claim or delegate. Bird~\cite{bird2024artists} documents a double bind between market relevance and artistic integrity. Kyi et al.~\cite{kyi2025governance} formalize artists' expectations around consent, credit, and compensation.

These accounts establish how AI affects creative identity. We follow the same practitioners through changing tools and norms. In HAI, identity and value negotiation form part of longitudinal agency partitioning. Identity work concerns maintaining a professional self. Agency partitioning identifies the workflow and claim boundaries that enact that identity. Its longitudinal form captures their repeated reopening across several years.

\section{Methodology}
We used a five-year longitudinal qualitative design from 2021 to 2025 to examine how digital painters' relationships with generative AI changed over time. We treated participants' accounts as situated interpretations of changing tools, values, and practices. Longitudinal qualitative research supports the analysis of change in identity, value, and practice~\cite{audulv2023time, saldana2003longitudinal, kjaerup2021longitudinal}. It also preserves the difference between accounts recorded at different moments~\cite{chalfen1987snapshot, dourish2006implications}. A stable annual interview guide preserved comparison while participants' own categories of AI, creativity, and agency shifted over time.

\subsection{Participants}

We recruited participants in 2021 through calls posted on Chinese social media platforms, including WeChat, Xiaohongshu, Bilibili, Baidu Tieba, and Weibo. Eligibility required participants to be at least 18 and have one year of digital painting experience. Participants also needed basic awareness of generative AI and willingness to complete annual interviews over five years. The sample focused on digital and illustrative practice. We excluded traditional painting, 3D modeling, video editing, and music production.

In this study, \textit{digital painters} refers to artists who primarily create two-dimensional visual works using digital tools, including Photoshop, Clip Studio Paint, and Procreate. Their practice spans illustration, concept design, and game or animation art. It often involves balancing artistic expression with client requirements and fast-paced production cycles. Seventeen participants were enrolled in 2021. The sample included 7 men and 10 women, aged 19 to 34 ($M=26.3$, $SD=4.37$). Six participants had postgraduate education, 10 had undergraduate education, and 1 had a high school education. Eleven identified as professional digital painters employed in game studios, publishing houses, or design firms. Six identified as hobbyists.

Interview availability varied across waves because of attrition and intermittent non-response. Available records by year were 17, 15, 15, 14, and 14 from 2021 to 2025. As detailed in Appendix~3 of the supplementary materials, 13 participants contributed data in all five waves. The remaining four contributed between one and four waves. We distinguish, where relevant, between claims about the baseline cohort, five-wave records, and participants with data available in a given wave. Table~\ref{tab:participants} summarizes baseline demographics. Appendix~3 of the supplementary materials details creative domains, client types, and professional status across waves.

This study was approved by the institutional review board at the authors' institution. Each participant received \textyen 40 for each completed annual interview, up to \textyen 200 over five years. Because interviews touched on sensitive professional topics, including layoffs, salary erosion, and intra-team conflict, participants could redact specific passages on request. Quotations are reported with permission and anonymized at the level of company names and identifying biographical details.

\begin{table}[htbp]
\centering
\caption{Summary of Participant Demographics (Baseline in 2021, $N=17$).}
\label{tab:participants}
\footnotesize
\setlength{\tabcolsep}{3pt}
\begin{tabular}{ccccccl}
\toprule
ID & Age & Gen. & Edu. & U/R & Yrs & Status \\
\midrule
1  & 22 & F & BA & U & 2  & Pro \\
2  & 24 & F & BA & U & 6  & Pro \\
3  & 22 & M & BA & R & 7  & Pro \\
4  & 19 & M & MA & U & 3  & Non-Pro \\
5  & 33 & F & MA & U & 11 & Pro \\
6  & 26 & F & MA & U & 8  & Pro \\
7  & 34 & M & BA & R & 8  & Pro \\
8  & 31 & M & MA & U & 7  & Pro \\
9  & 29 & F & BA & R & 8  & Pro \\
10 & 26 & M & BA & U & 8  & Non-Pro $\to$ Pro \\
11 & 26 & F & HS & R & 7  & Pro \\
12 & 25 & F & MA & U & 7  & Non-Pro \\
13 & 33 & M & BA & U & 8  & Non-Pro \\
14 & 22 & F & BA & R & 7  & Non-Pro \\
15 & 27 & M & BA & R & 9  & Pro \\
16 & 24 & F & BA & R & 11 & Pro (att.) \\
17 & 24 & F & MA & R & 6  & Non-Pro (att.) \\
\bottomrule
\end{tabular}\\[2pt]
{\scriptsize Edu. = BA Bachelor, MA Master, HS High School. U/R = Urban/Rural. ``att.'' = attrited before study end.}
\end{table}

\subsection{Procedure}

Each year, participants completed one 60 to 90 minute semi-structured interview online via Tencent Meeting. Two researchers jointly conducted every annual interview. The guide remained unchanged across all five waves. It contained four blocks covering background, experience with generative AI, attitudes toward generative AI, and forward-looking reflections. Questions prompted participants to discuss the past year's tools, workflows, and perceived professional impacts. The full protocol appears in Appendix~1 of the supplementary materials.

Sessions were audio-recorded with consent. Interviews were conducted in Mandarin Chinese. Transcripts were produced verbatim within 48 hours and cross-checked by two researchers. Quotes are translated from Chinese by bilingual researchers. Analytically important terms, including ``creation,'' ``originality,'' ``AI art,'' and dismissive descriptors, were discussed to preserve participant stance and local idiom. Analytically central quotations were checked against the Mandarin originals by at least two bilingual researchers.

\textbf{Anchoring the 2021 baseline.} ``Generative AI'' in 2021 referred to a different ecosystem from today's prompt-controllable diffusion systems. Participants' 2021 awareness was shaped mainly by GAN-based style-transfer and avatar generators, AI retouching features such as Photoshop's Neural Filters, and early image-generation discussions in online art and ACG communities~\cite{discodiffusion}. These differed from the widely accessible prompt-controllable systems and model-specific discourse that circulated more broadly after 2022~\cite{novelai}. We treat this technological and conceptual drift as part of the phenomenon. Anchoring the protocol around participants' lived practice let us follow shifts through participants' descriptions of tools, stages, and workflows each year.

\subsection{Data Analysis}

\subsubsection{Codebook-Based Thematic Analysis}
We used NVivo to conduct codebook-based thematic analysis. Two researchers independently coded transcripts from two participants across all five waves to develop an initial codebook. The coding unit was a meaning unit, such as a response, follow-up exchange, or coherent workflow account. Percent agreement on this calibration set exceeded 80\%. The percentage assessed consistent application of the initial codebook during coding calibration. It was not used to establish theme validity. After calibration, the first author coded the remaining transcripts. The team reviewed uncertain and newly emergent codes in regular meetings, resolved differences through discussion, and documented decisions in analytic memos. New codes were added at each wave, with revisions reviewed by the broader team.

As a complement to thematic coding, we maintained lightweight lexical memos about recurrent descriptors participants used for AI imagery and AI-assisted work. These included dismissive terms such as \textit{ugly}, \textit{trash}, \textit{soulless}, \textit{collage}, and \textit{gibberish}, along with later pragmatic terms around efficiency, reference, and workflow. These memos helped the team notice shifts in tone across waves. Interview availability, transcript length, and participants' vocabularies changed over time, so the memos serve as contextual support. The main evidence comes from year-stamped quotations and longitudinal case comparisons.

For the agency partitioning analysis, we marked passages where participants assigned a role, task, responsibility, or claim to either themselves, AI, peers, clients, or platforms. Relevant passages included delegated workflow stages, human-reserved stages, authorship claims, credit concerns, labor-value judgments, and peer or client legitimacy judgments. We retained the year of each passage during coding. This allowed us to distinguish boundaries that remained stable from boundaries reopened by later tool use, workplace pressure, or community debate.

\subsubsection{Longitudinal Case Analysis}
Building on the thematic framework, we created longitudinal case files and synchronously coded each participant's available annual records as an evolving narrative. Each interview remained a year-specific account, so later reinterpretation did not replace earlier accounts. Following longitudinal qualitative research~\cite{kjaerup2021longitudinal, lewis2007analysing, vogl2018developing}, we examined \textit{narrative developments}, \textit{reinterpretations}, and \textit{stability}. For example, we compared cases where a 2023 description of efficiency was later reframed as exploitation in 2025. We kept missing waves as missing data. Claims about change are restricted to participants and years with available records. Codes were aggregated into themes and subthemes that the broader team reviewed for coherence. Participant context tables, codebook documentation, analytic memos, and verbatim year-stamped quotations formed the analysis record. By the later waves, several themes had become recurrent. We treated the 2025 interviews as an opportunity to examine reinterpretation and consolidation.

\section{Findings}
Our findings show three recurring but non-uniform patterns of agency partitioning over time. These were protective resistance from 2021 to 2022, pragmatic task delegation from 2022 to 2024, and reflective agency repartitioning for some participants by 2025. Divergent paths included reframing, procedural compromise, continued refusal, difficulty locating a human role, and labor-related precarity.

Agency partitioning was visible when participants assigned responsibility to a specific part of creative work. Evidence included delegated stages, human-only authorship claims, final delivery responsibility, and aesthetic, economic, peer, or client legitimacy boundaries. Peer norms and emotional climates shaped delegation, while copyright and authorship remained recurring limits. Appendix~4 of the supplementary materials summarizes phase-level evidence across waves.

The three labels are analytic anchors rather than exclusive participant types. A participant could protect one boundary while delegating another in the same year. A professional could use AI for backgrounds under deadline pressure while rejecting AI authorship. A hobbyist could refuse AI in finished work while experimenting with prompts or references. The phases organize allocations across tasks, claims, and social settings while preserving divergent cases.

\subsection{Longitudinal Trajectory of Agency Partitioning}
\label{subsec:trajectory}

\subsubsection{Protective Resistance from 2021 to 2022}

Between 2021 and 2022, many participants moved from tentative curiosity toward protective resistance. In 2021, many described themselves as ``testing the waters.'' P3 recalled, \textit{``I played with it for fun, just to see what it could do. At that time it felt like a toy, not something that could threaten me'' (2021).} By 2022, as AI-generated images circulated widely on social media and art forums, the tone shifted in available accounts. Participants frequently dismissed outputs as ``garbage,'' ``worthless collages,'' or ``devoid of artistic sensibility.'' P1, who had spoken of curiosity in 2021, concluded, \textit{``I tried again this year, and honestly, it became worse. Everyone is posting AI art everywhere, and to me they all look the same, shallow, soulless'' (2022).}

A common criticism concerned aesthetic shallowness. P9 added, \textit{``Maybe they look polished, but they have no warmth. Real art always carries traces of the artist's struggle, and these images erase that struggle'' (2022).} A second critique targeted creativity itself. P12 asked, \textit{``Can patching things together really count as creativity?'' (2022).} P5 elaborated. \textit{``The real problem isn't quality. It's that AI doesn't actually observe anything. When I draw a forest, I'm drawing from somewhere specific I've been, the angle of the light, the weight of the air. AI doesn't observe any of that. It derives what a forest `statistically looks like' and generates it. That's not creation, that's retrieval. Watching people call that `art'... I can't accept it.'' (P5, 2022).}
P5's 2022 stance can be read against their more cautious curiosity one year earlier. \textit{``I downloaded a couple of those apps and tried them for an afternoon. The results were honestly pretty bad. You could tell it was just grabbing pieces from its training data and cobbling them together. You type in `warrior' and it gives you something vaguely warrior-shaped, but the proportions are off, the logic is off. It just doesn't know what you want. It's guessing.'' (2021).} The shift from technical critique in 2021, where AI did not \textit{know}, to a sharper denial of perceptual agency in 2022, where AI did not \textit{observe}, shows how resistance became a way to withhold creative capacities from the system.

Authorship and copyright concerns were part of the same boundary work. P14 asked, \textit{``If people can't tell the difference, what does my signature mean anymore? My name used to guarantee originality. Now it might not'' (2022).} P7 worried about training data, \textit{``If someone takes my work to train a dataset, do I still own the outputs?'' (2022).} P2 framed the issue as livelihood theft, \textit{``If AI can borrow from my drawings without asking, it's not just unfair. It's stealing my future opportunities'' (2022).} Early resistance therefore marked creative capacities, signatures, and ownership claims that participants did not want absorbed into an opaque agentic system.

\subsubsection{Pragmatic Task Delegation from 2022 to 2024}

Between 2022 and 2024, several accounts moved from rejecting AI as a creative actor to delegating local workflow tasks. Dismissive descriptions remained salient in 2022, while later accounts emphasized utility, visual quality, and creative usefulness. Participants used AI for references, color schemes, backgrounds, rough sketches, and client-facing drafts. Final selection, correction, signature style, and delivery responsibility remained human. Professionals linked these uses to deadlines, team workflows, and client expectations within speed-oriented pipelines. Hobbyists retained more freedom to experiment or refuse. These bounded placements were later interpreted as efficiency, fatigue, or labor devaluation.

\textbf{Reappraising artistic quality.} By 2023, several participants acknowledged that AI outputs had improved while maintaining doubts about creativity. P6 reflected, \textit{``I saved many AI-generated comic characters on my phone... their color schemes look really appealing, and I can use them as references when drawing'' (2023).} P11 acknowledged the increasing polish, \textit{``If you don't explicitly disclose it, many AI artworks are already hard to distinguish from human-made pieces'' (2023).}

\textbf{Efficiency as adaptation.} For professionals, the more important factor was adaptation to production pipelines. P10's three-year arc shows a slow conversion from peer pressure to operational acceptance. In 2022, they described feeling cornered because \textit{``my colleagues kept sending AI drafts in our group chat. If I ignored them, I felt out of touch''}. By 2024, they said, \textit{``Even if I don't like AI art, I have no choice. My colleagues are already using it. What takes me ten days to draw, AI can finish in a minute. I still stubbornly believe my work looks better, but that doesn't matter'' (2024).} P2 echoed this pragmatic use, \textit{``Using AI to generate background images is extremely convenient. Feed it a dataset and it's done in a few minutes'' (2024).} The aesthetic objections raised in 2022 persisted as private misgivings. Deadlines and team workflows displaced them in everyday practice. Painters could reject AI as an author while accepting it as a production shortcut for backgrounds, references, and other bounded tasks.

\textbf{AI as ideation scaffold.} Participants also incorporated AI into early conceptualization. P15 described, \textit{``Many times I just tell the AI the general idea, let it generate a rough sketch, and then use that sketch to communicate with the client'' (2024).} Participants often treated outputs as raw material for later human judgment. This placement allowed them to delegate ideation or reference generation while preserving human selection, correction, and final claim-making.

Pragmatic delegation coexisted with earlier resistance by moving AI into workflows while keeping authorship, responsibility, taste evaluation, and the final creative claim human. The boundary shifted from broad rejection to task-level placement. \textbf{Professional status as contextual condition.} Delegation was unevenly distributed. Commercial pipelines appeared to encourage earlier or more frequent use among baseline professionals P1, P2, P10, P11, and P15 despite personal ambivalence. Baseline hobbyists P4, P12, P13, and P14 faced fewer market constraints and retained more freedom to remain skeptical or selective through 2024.

\subsubsection{Divergent Endpoints and Reflective Agency Repartitioning in 2025}

By 2025, enthusiasm had cooled and critical voices resurfaced. The focus of critique had also shifted. Participants with 2025 records grappled with what AI meant for livelihoods, artistic values, and identity. Some described explicit strategies for repartitioning human and AI roles. Others described fatigue, precarity, continued refusal, or difficulty locating a remaining human role.

\textbf{Labor and industry pressures.} Efficiency gains no longer inspired only optimism. They also brought labor and compensation tensions into view. Five of the 14 participants with 2025 records reported shrinking salaries, fewer opportunities, or unemployment. P8 offers one illustrative arc. In 2022, P8 was a game-art practitioner at a large company and felt their role was insulated from AI's reach. \textit{``AI is fine for amateurs, but our studio runs on composition, mood, narrative consistency. Things AI can't reach. The studio knows this'' (2022).} After leaving stable company work, they had revised the frame by 2025. \textit{``After I left, I finally understood the math. AI produces in three minutes what took me a month to do. When employers run that calculation, salary is the first thing that moves. When I was still there, I was making around 7,000 RMB. I know people still in the industry getting offered half that now, and still expected to use AI to make up the difference.'' (2025).} P2 noted the irony of unrewarded productivity.
\textit{``It's kind of counterintuitive. Before, one concept drawing a week was considered fast at my company. Now with AI I can turn out three or four a day. But the money? Pretty much the same. And because I deliver faster, clients feel like they can ask for more revision rounds, more options, `you're fast anyway.' My output has multiplied, but the value of each piece has gone down.'' (P2, 2025).}
P2, an illustrator who shifted from a small company toward private concept-design work, moved from ethics-first rejection in 2022 to background delegation in 2024. By 2025, efficiency had become a source of possible labor devaluation, making economic value a boundary around AI delegation.

P4 was an urban Master's student in design whose early reactions were the cohort's most vehement, including ``visual gibberish'' in 2022. In 2025, P4 described a more existential uncertainty. \textit{``I keep asking myself what I actually bring anymore. Composition? AI can do that. Color, definitely. Line quality is almost there. Character design, visual storytelling. I'm not sure anymore. I spent years grinding these specific skills, and then I open one of these tools and it's done in fifteen seconds. I'm not even angry. I just don't know where I fit.'' (2025).} This account marks an endpoint where repartitioning was not easily achieved. For some participants, the acceleration valued in 2023 and 2024 became precarity that made human labor feel easier to discount.

\textbf{Aesthetic fatigue.} Six of the 14 participants with 2025 records reported that AI's technical polish had produced a sense of monotony. P13 explained, \textit{``today's AI drawings are no longer obviously AI, but if you look closely you can still tell. There are no traces of handcraft, too perfect in a way that feels unsettling'' (2025).} P9's 2025 weariness contrasted with their 2022 critique. Three years earlier, they had charged AI with lacking \textit{``warmth''}. In 2025, they echoed that critique from the position of a more saturated user. \textit{``Everywhere I look, platforms, forums, galleries, it's flooded with AI works. After a while, they all start to look the same'' (2025).} The concern persisted, but its emotional tone shifted from indignation toward fatigue.

\textbf{Reclaiming identity.} By 2025, some participants described hybrid workflows as a way to repartition agency. Two of the 14 participants with 2025 records stated they would limit AI to early ideation to avoid ``AI-ization'' of their work. Others used hybrid workflows to turn the human and AI boundary into a sequence of labor steps. P14, a non-professional participant, described their strategy.
\textit{``I figured out this way of working. The sketch I never skip, that part is always by hand. Then I let AI handle the base colors, because coloring is where I always got stuck forever. After that I go back and fix everything that looks off. There's always something. The final pass and all the details are mine. Four steps total. It sounds complicated, but it's the only way I feel like the drawing is still mine.'' (P14, 2025).}
P14 was a non-professional rural participant outside formal client work whose authorship concerns were primarily symbolic and craft oriented. The four-step \textit{workflow partition} reserved three steps for the human and enforced the boundary through their sequence. It turned P14's 2022 signature concern into a concrete authorship and workflow boundary in 2025. P6 offered a more symbolic boundary. \textit{``I use AI now, for backgrounds, color references, sometimes rough layout. But the character's face, I don't let it touch that. The face is where everything lives, the expression, the personality. If I let AI design the face, it starts to feel like I'm just approving someone else's work instead of drawing.'' (2025).} Reserving the face marks an aesthetic-signature boundary where AI assistance should stop. Selective integration made the remaining human contribution visible and meaningful. P2 reframed, P14 redesigned their workflow, and P4 deepened rejection into existential reckoning. These cases show distinct pathways within the recurring pattern.

\subsection{Conditions That Reopened or Preserved Agency Boundaries}
\label{subsec:cross-cutting}

The phases also reflected changing peer norms, emotional responses, and ethical constraints. Peer influence moved from pressure toward normalization, polarization, and withdrawal. Affect shifted from curiosity toward offense, awe, and fatigue. Copyright and authorship persisted while practical use changed.

\subsubsection{From Peer Pressure to Normalization and Polarization}

Peer and community dynamics shaped how painters interpreted AI. In 2022, peer influence was experienced as pressure. P10 explained, \textit{``Even if I didn't want to use it, my colleagues kept sending AI drafts in our group chat. If I ignored them, I felt out of touch'' (2022).} P3 echoed this concern, \textit{``I didn't trust it, but everyone around me was testing it. I was worried I'd be left behind'' (2022).}

By 2023 and 2024, peer influence appeared to shift from subtle pressure toward social normalization. P11 observed, \textit{``Last year people laughed at AI drawings. Now, in our group, everyone posts them casually'' (2023).} P6 noted that client expectations had begun to shift. \textit{``A client recently asked me, almost casually, whether we use AI in the concept sketch phase. Not as a concern, just like it was already the standard. We were still debating it internally, and clients were already treating it as a baseline.'' (2024).} By 2025, four of the 14 participants with 2025 records reported community fragmentation along AI-rejecting and AI-embracing lines. P14 captured the experience most explicitly, \textit{``Some of my friends insist that real artists must reject AI completely, while others post AI drafts every day. Our conversations sometimes turn into arguments. I feel caught in between'' (2025).} P9 described withdrawal from the debate. \textit{``I ended up muting three or four group chats. Every morning it was people showing off their latest outputs, which model, how many runs. Then by afternoon it turned into someone ranting that AI is destroying the industry. I was tired of both sides. Same loop, nobody listening. I just wanted to draw my own things.'' (2025).}

These accounts show how peers and clients shaped which delegations became professionally useful or morally acceptable. They also show that fatigue, polarization, and withdrawal affected whether participants experimented, resisted, or imposed limits.

\subsubsection{From Curiosity and Offense to Awe and Fatigue}

Participants' accounts also traced shifts in emotional responses to generative AI. In 2021, curiosity dominated. P3 recalled treating AI \textit{``like a toy. I typed in silly prompts just to see what came out, and it was funny more than anything else'' (2021).} In 2022, affect shifted toward frustration and offense. P4 exclaimed, \textit{``It's everywhere now, and it makes me mad'' (2022).} P12 added, \textit{``When people call these collages `creative,' it feels like an insult to years of training'' (2022).}

By 2023 and 2024, surprise and awe surfaced. P6 admitted, \textit{``I was shocked. The images suddenly looked good. Sometimes I even felt inspired'' (2023).} P11 described exhilaration, \textit{``The first time I saw an AI piece that truly amazed me, I felt both scared and thrilled'' (2024).} By 2025, the emotional tone had cooled into fatigue. P13 said, \textit{``I don't feel shocked anymore. AI works are everywhere, and they all feel a bit too perfect. After a while it's numbing'' (2025).} P14 summarized the arc, \textit{``At first I was excited, then I felt overwhelmed, and now I just want to rethink what role it should play in my work'' (2025).} Affect mattered because participants linked it to whether they entered, avoided, or limited collaboration with AI.

\subsubsection{Ethical Boundaries Persisted as Use Changed}
\label{subsubsec:ethical-anchor}

Copyright, data provenance, and authorship followed a steadier pattern across available accounts than aesthetic and pragmatic evaluations. These concerns appeared when participants discussed training data, client deliverables, and hybrid authorship.

In 2022, when many participants were aesthetically dismissive of AI, copyright surfaced as a reason for moral and artistic rejection. P7 worried, \textit{``If someone takes my work to train a dataset, do I still own the outputs?'' (2022).} P2 framed the issue as livelihood theft, \textit{``If AI can borrow from my drawings without asking, it's stealing my future opportunities'' (2022).}

By 2023 and 2024, when several participants described pragmatic AI use in client or production workflows, copyright concerns moved into new contexts. Participants asked who owned AI-assisted client deliverables, whether artists could refuse data scraping, and how hybrid human-agent works should be credited. Without stable answers, copyright functioned as a background standard against which participants evaluated new AI capabilities.

By 2025, copyright and authorship had become difficult to separate from identity reconstruction. P14's four-step hybrid workflow was also an authorship-preservation strategy. By leaving certain steps ``purely mine,'' painters marked the boundaries of their legal and moral claims to the work. P6's claim that ``the face is where everything lives'' (Sec.~\ref{subsec:trajectory}) makes this hybrid arrangement visible as ethical self-positioning. Practical attitudes toward AI tools changed substantially, but ownership and authorship remained limits on what creators felt comfortable delegating.

\section{Discussion}
Chinese digital painters negotiated creative agency repeatedly as generative AI changed. \textit{RQ1.} As prompt-controllable systems became professionally relevant after 2022, participants described protective resistance, pragmatic task delegation, and reflective repartitioning of what should remain human. \textit{RQ2.} Aesthetic judgments and practical uses changed substantially, while copyright, authorship, and creative labor remained boundaries around delegation (Sec.~\ref{subsubsec:ethical-anchor}). \textit{RQ3.} Adaptation was community mediated. Peer norms and emotional climates shaped whether participants reframed AI, redesigned workflows, or resisted its place in their identity (Sec.~\ref{subsec:cross-cutting}).

\subsection{Longitudinal Agency Partitioning}

Digital painters' accounts moved beyond a linear path from rejection to acceptance. Agency partitioning identifies how tasks, responsibilities, values, and claims are allocated at a given point. Agency repartitioning identifies a later revision. Longitudinal agency partitioning captures repeated maintenance, reopening, and revision. Agency refers to participants' attributed responsibility and control. The six overlapping boundaries involved workflow delegation, authorship, aesthetic signature, moral or legal responsibility, economic value, and peer legitimacy.

Resistance in 2021 and 2022 questioned whether AI could participate meaningfully in creative expression~\cite{kirova2023ethics, lima2025public}. During 2023 and 2024, several participants delegated reference generation, background drafting, and client communication sketches. They kept final judgment, correction, and authorship human. By 2025, some accounts shifted from whether to use AI toward where its agency should end. The Findings and Appendix~4 of the supplementary materials show boundary work around final judgment, responsibility, economic value, and signature style. P4 marks a different endpoint where repartitioning was not easily achieved.

Our framework draws on artists' double binds~\cite{bird2024artists}, AI-related identity work~\cite{munoz2023identity}, and the consent, credit, and compensation framework~\cite{kyi2025governance}. It adds a temporal account of how similar concerns are revisited under changing social, technical, and economic conditions. Consent was especially visible in 2022. Credit became salient as hybrid workflows proliferated. Compensation concerns intensified late, as participants described higher output without comparable pay growth. These shifts suggest that long-term creative interaction can reopen agency boundaries that once seemed settled.

Identity and value negotiation can be read as one practical face of a broader HAI concern, ongoing agency attribution under uncertainty~\cite{glikson2020human, meng2026living}. The advisor, collaborator, and driver typology reviewed in Sec.~2.2 treats agency attribution as a property of interaction design. Our findings suggest that attribution can also be revisited across time. The craft-level question is who has agency over what, under which social conditions, and for how long.

Longitudinal HAI research can examine the boundary that a participant maintains, reopens, or revises across encounters. Using the same interview guide across waves anchors comparison while participants' meanings of AI and creativity continue to change. Year-stamped accounts and participant case files preserve stability, reinterpretation, and missing waves~\cite{audulv2023time, kjaerup2021longitudinal, vogl2018developing}. This approach distinguishes a recurring agency boundary from a temporary response to a new capability or social pressure.

\subsection{Peer and Affective Conditions}
\label{subsec:resonance}

Peer influence and emotional climate shaped agency partitioning. As peers embraced AI, clients and communities increasingly treated its use as expected. Participants described refusal as a potential source of marginalization~\cite{vannoy2010social, woodruff2024knowledge, zhang2024confrontation}. Anxiety appeared early, excitement and awe surfaced in the middle waves, and fatigue and ambivalence became salient later~\cite{cetinic2022understanding, johnston2024understanding}. These conditions form a descriptive account of agency partitioning.

Across cases, peer circulation could make refusal professionally risky, community saturation could make AI use exhausting, and polarization could encourage compromise or withdrawal. For HAI, creative agents may be interpreted through system interaction and peer settings where outputs are disclosed, attributed, circulated, and debated.

\subsection{The Chinese Digital-Painting Context}

The Chinese digital-painting industry shaped these patterns. \textit{Industrialized illustration} for mobile games, online novels (\textit{wangwen}), web comics (\textit{guoman}), and short-video animation makes production efficiency a visible business metric~\cite{lu2025research, zhang2020platform, zhang2023labor}. This clarifies the labor-economic interpretation of efficiency. \textit{Platform-mediated peer ecologies}, including WeChat, Bilibili, Xiaohongshu, Lofter, and Tieba, circulate output styles, copyright news, tutorials, and moral judgments~\cite{scheen2019boredom}. Chinese AIGC court rulings in 2023 and 2024 formed part of the context in which hybrid workflows became common~\cite{yang2024rethinking}.

Agency partitioning felt urgent because pipelines rewarded speed, platforms amplified peer comparison, and copyright debates supplied vocabulary for claims over hybrid work.

\subsection{Design Considerations}

The design opportunity is to make agency boundaries visible at the level where painters described them. These levels included stages, layers, versions, public posts, client deliverables, and authorship claims. Designers can treat boundary setting as part of the creative workflow and revisit it again at export~\cite{liu2025supporting}.

\textbf{Agency-boundary controls.} Prior HCI research shows that advisor, collaborator, and driver roles distribute control differently~\cite{schecter2025role, hu2025designing, lawton2023tool}. Participants made comparable boundary decisions at the level of concrete tasks. They preserved human zones such as sketching, final correction, character faces, and signature style. Creative AI tools could let users mark workflow stages as human-only, AI-assisted, or AI-generated. Similar controls could operate through layers, masks, version histories, or review checkpoints. A creator might lock a face layer as human-only, mark a background as AI-assisted, and keep final correction as a human pass. These markings should remain revisable across projects and career stages.

\textbf{Provenance and authorship scaffolds.} Prior HCI work identifies consent, credit, compensation, ownership, and transparency as central governance concerns~\cite{kyi2025governance, shelby2024generative, lovato2024foregrounding}. These concerns recurred even when participants pragmatically adopted AI. Systems could document prompts, source references, generated intermediates, manual edits, and final human interventions. Such records could support credit, disclosure, and authorship claims over time. Provenance could also separate private workflow records from public-facing disclosure. This distinction matters because clients, employers, peers, and platforms hold different expectations. Provenance logs should be treated as accountability infrastructure. Legal ownership, consent, compensation, and labor value remain institutional questions.

\textbf{Layered collaboration and community-facing norms.} Creative AI use is shaped by social influence, community judgment, and human-centered collaboration~\cite{bird2024artists, woodruff2024knowledge, zhang2024confrontation, rezwana2025human}. Our findings locate these pressures in peers, clients, and platforms. The same practice could be read as efficient, dishonest, normal, or exhausting. Systems could support movement among advisory, assistive, and co-creative modes while preserving protected human zones~\cite{chen2026between, meng2026balancing, meng2026personalized}. Platforms could provide revisable disclosure tags, provenance displays, and attribution norms that make disagreements over AI use more accountable. Community-facing designs could let artists specify AI's role in a post or portfolio item through granular labels. These norms should remain sensitive to communities with different values, labor conditions, and risk tolerance~\cite{zeng2025parental, chen2026not}.

Our results suggest that adoption is intertwined with value systems, professional positioning, and agency attribution. Long-term creative human-agent interaction points toward a view of agency that is temporal, socially mediated, and morally bounded.

\subsection{Limitations and Future Work}
\label{sec:limitations}

\textbf{Sample and attrition.} Seventeen baseline participants and 13 five-wave records form a modest, self-selecting cohort. We trace plausible longitudinal patterns within this cohort. Broader samples could examine whether professionals delegate tasks to AI more rapidly than hobbyists.

\textbf{Cultural, domain, and conceptual specificity.} The Chinese digital-painting context shapes our findings. Meanings of ``AI-generated art,'' ``originality,'' and ``creativity'' also shifted. Cross-cultural and cross-domain work~\cite{meng202652, meng2026creating} with adaptive probes could clarify which themes generalize.

\textbf{Recall, retrospective bias, and repeated interviewing.} Participants may re-narrate earlier waves through later experiences, while repeated interviews may shape accounts of continuity and change. We retained each interview as a situated annual account before comparison. Real-time diaries could provide finer-grained affective records.

\textbf{Disentangling tool affordances from identity work.} We did not isolate prompt engineering, controllable outputs, or video generation from sociocultural identity work. Targeted system evaluations~\cite{ma2026can, zhang2025slideaudit, luo2025s} could complement longitudinal interviews.

\section{Conclusion}
This five-year longitudinal interview study of 17 Chinese digital painters found recurring but non-uniform patterns of protective resistance, pragmatic task delegation, and reflective agency repartitioning. Copyright, authorship, and creative labor remained recurring boundaries on delegation, while peer norms and emotional climates shaped whether AI use felt useful, acceptable, or exhausting. We contribute longitudinal agency partitioning as a lens for studying creative human-agent interaction over time. Grounded in this cohort and context, it traces repeated negotiation through workflow practice, authorship claims, peer circulation, and labor value.

\section*{Acknowledgments of the Use of AI}
Generative AI provided only language support for grammar and readability, plus LaTeX formatting assistance for tables and appendices. It was not used for data collection, transcript coding, theme generation, analysis, interview materials, or participant data. The authors remain responsible for the integrity of the data, analysis, and claims.

\bibliographystyle{ACM-Reference-Format}
\bibliography{ref}

@article{audulv2023time,
  title={Time and change: a typology for presenting research findings in qualitative longitudinal research},
  author={Audulv, {\AA}sa and Westergren, Thomas and Ludvigsen, Mette Spliid and Pedersen, Mona Kyndi and Fegran, Liv and Hall, Elisabeth OC and Aagaard, Hanne and Robstad, Nastasja and Kneck, {\AA}sa},
  journal={BMC Medical Research Methodology},
  volume={23},
  number={1},
  pages={284},
  year={2023},
  publisher={Springer}
}

@book{saldana2003longitudinal,
  title={Longitudinal qualitative research: Analyzing change through time},
  author={Salda{\~n}a, Johnny},
  year={2003},
  publisher={Bloomsbury Publishing PLC}
}

@article{chalfen1987snapshot,
  title={Snapshot versions of life},
  author={Chalfen, Richard},
  year={1987},
  publisher={Bowling Green State University Popular Press Bowling Green, OH}
}

@inproceedings{dourish2006implications,
  title={Implications for design},
  author={Dourish, Paul},
  booktitle={Proceedings of the SIGCHI conference on Human Factors in computing systems},
  pages={541--550},
  year={2006}
}

@book{kjaerup2021longitudinal,
  title={Longitudinal studies in HCI research: a review of CHI publications from 1982--2019},
  author={Kj{\ae}rup, Maria and Skov, Mikael B and Nielsen, Peter Axel and Kjeldskov, Jesper and Gerken, Jens and Reiterer, Harald},
  year={2021},
  publisher={Springer}
}

@article{lewis2007analysing,
  title={Analysing qualitative longitudinal research in evaluations},
  author={Lewis, Jane},
  journal={Social Policy and Society},
  volume={6},
  number={4},
  pages={545--556},
  year={2007},
  publisher={Cambridge University Press}
}

@article{vogl2018developing,
  title={Developing an analytical framework for multiple perspective, qualitative longitudinal interviews (MPQLI)},
  author={Vogl, Susanne and Zartler, Ulrike and Schmidt, Eva-Maria and Rieder, Irene},
  journal={International Journal of Social Research Methodology},
  volume={21},
  number={2},
  pages={177--190},
  year={2018},
  publisher={Taylor \& Francis}
}

@inproceedings{bird2024artists,
  title={Artists and AI: Creative Interactions and Tensions},
  author={Bird, Charlotte},
  booktitle={Extended Abstracts of the CHI Conference on Human Factors in Computing Systems},
  pages={1--6},
  year={2024}
}

@inproceedings{inie2023designing,
  title={Designing participatory ai: Creative professionals’ worries and expectations about generative ai},
  author={Inie, Nanna and Falk, Jeanette and Tanimoto, Steve},
  booktitle={Extended Abstracts of the 2023 CHI Conference on Human Factors in Computing Systems},
  pages={1--8},
  year={2023}
}

@inproceedings{page2025creative,
  title={Creative Reflections on Image-Making with Artificial Intelligence: Interactions with a Provocative'Camera'},
  author={Page, Rowan and See, Jian Shin},
  booktitle={Proceedings of the 2025 CHI Conference on Human Factors in Computing Systems},
  pages={1--16},
  year={2025}
}

@inproceedings{he2023exploring,
  title={Exploring designers’ perceptions and practices of collaborating with generative AI as a Co-creative agent in a multi-stakeholder design process: take the domain of avatar design as an example},
  author={He, Qingyang and Zheng, Weicheng and Bao, Hanxi and Chen, Ruiqi and Tong, Xin},
  booktitle={Proceedings of the Eleventh International Symposium of Chinese CHI},
  pages={596--613},
  year={2023}
}

@inproceedings{chung2022artistic,
  title={Artistic user expressions in AI-powered creativity support tools},
  author={Chung, John Joon Young},
  booktitle={Adjunct Proceedings of the 35th Annual ACM Symposium on User Interface Software and Technology},
  pages={1--4},
  year={2022}
}

@inproceedings{lu2025designing,
  title={Designing and Developing User Interfaces with AI: Advancing Tools, Workflows, and Practices},
  author={Lu, Yuwen and Jiang, Yue and Knearem, Tiffany and Kliman-Silver, Clara E and Lutteroth, Christof and Nichols, Jeffrey and Stuerzlinger, Wolfgang},
  booktitle={Proceedings of the Extended Abstracts of the CHI Conference on Human Factors in Computing Systems},
  pages={1--7},
  year={2025}
}

@inproceedings{chung2021intersection,
  title={The intersection of users, roles, interactions, and technologies in creativity support tools},
  author={Chung, John Joon Young and He, Shiqing and Adar, Eytan},
  booktitle={Proceedings of the 2021 ACM Designing Interactive Systems Conference},
  pages={1817--1833},
  year={2021}
}

@inproceedings{kawakami2024impact,
  title={The impact of generative ai on artists},
  author={Kawakami, Reishiro and Venkatagiri, Sukrit},
  booktitle={Proceedings of the 16th Conference on Creativity \& Cognition},
  pages={79--82},
  year={2024}
}

@inproceedings{sikorski2025attitudes,
  title={On the Attitudes of GameDev Industry Artists towards GenAI. Preliminary Results},
  author={Sikorski, {\L}ukasz and Matulewski, Jacek and Czerwonka, Ma{\l}gorzata},
  booktitle={Proceedings of the 2025 Computers and People Research Conference},
  pages={1--6},
  year={2025}
}

@inproceedings{porquet2025copying,
  title={Copying style, Extracting value: Illustrators' Perception of AI Style Transfer and its Impact on Creative Labor},
  author={Porquet, Julien and Wang, Sitong and Chilton, Lydia B},
  booktitle={Proceedings of the 2025 CHI Conference on Human Factors in Computing Systems},
  pages={1--16},
  year={2025}
}

@inproceedings{didion2024did,
  title={Who did it? How User Agency is influenced by Visual Properties of Generated Images},
  author={Didion, Johanna K and Wolski, Krzysztof and Wittchen, Dennis and Coyle, David and Leimk{\"u}hler, Thomas and Strohmeier, Paul},
  booktitle={Proceedings of the 37th Annual ACM Symposium on User Interface Software and Technology},
  pages={1--17},
  year={2024}
}

@inproceedings{schecter2025role,
  title={How the Role of Generative AI Shapes Perceptions of Value in Human-AI Collaborative Work},
  author={Schecter, Aaron and Richardson, Benjamin},
  booktitle={Proceedings of the 2025 CHI Conference on Human Factors in Computing Systems},
  pages={1--15},
  year={2025}
}

@inproceedings{hu2025designing,
  title={Designing Interactions with Generative AI for Art and Creativity: A Systematic Review and Taxonomy},
  author={Hu, Xi and Xing, Yiwen and Cai, Xudong and Zhao, Yihang and Cook, Michael and Borgo, Rita and Neate, Timothy},
  booktitle={Proceedings of the 2025 ACM Designing Interactive Systems Conference},
  pages={1126--1155},
  year={2025}
}

@inproceedings{kyi2025governance,
  title={Governance of Generative AI in Creative Work: Consent, Credit, Compensation, and Beyond},
  author={Kyi, Lin and Mahuli, Amruta and Silberman, M Six and Binns, Reuben and Zhao, Jun and Biega, Asia J},
  booktitle={Proceedings of the 2025 CHI Conference on Human Factors in Computing Systems},
  pages={1--16},
  year={2025}
}

@inproceedings{shelby2024generative,
  title={Generative AI in creative practice: ML-artist folk theories of T2I use, harm, and harm-reduction},
  author={Shelby, Renee and Rismani, Shalaleh and Rostamzadeh, Negar},
  booktitle={Proceedings of the 2024 CHI Conference on Human Factors in Computing Systems},
  pages={1--17},
  year={2024}
}

@inproceedings{johnston2024understanding,
  title={Understanding visual artists’ values and attitudes towards collaboration, technology, and AI},
  author={Johnston, Hannah and Thue, David},
  booktitle={Proceedings of the 50th Graphics Interface Conference},
  pages={1--9},
  year={2024}
}

@inproceedings{lawton2023tool,
  title={When is a tool a tool? user perceptions of system agency in human--ai co-creative drawing},
  author={Lawton, Tomas and Grace, Kazjon and Ibarrola, Francisco J},
  booktitle={Proceedings of the 2023 ACM Designing Interactive Systems Conference},
  pages={1978--1996},
  year={2023}
}

@inproceedings{xu2024application,
  title={The Application of Artificial Intelligence in the Creation of Four Grid Painting and Product Design},
  author={Xu, Linlin and Cheng, Pengfei},
  booktitle={Proceedings of the 2024 International Conference on Artificial Intelligence, Digital Media Technology and Interaction Design},
  pages={536--543},
  year={2024}
}

@article{canet2022dream,
  title={Dream Painter: Exploring creative possibilities of AI-aided speech-to-image synthesis in the interactive art context},
  author={Canet Sola, Mar and Guljajeva, Varvara},
  journal={Proceedings of the ACM on Computer Graphics and Interactive Techniques},
  volume={5},
  number={4},
  pages={1--11},
  year={2022},
  publisher={ACM New York, NY, USA}
}

@inproceedings{darabipourshiraz2025ai,
  title={AI DoodleLab: Fostering Middle School Students’ AI Literacy through Project-Based Creative Drawing},
  author={Darabipourshiraz, Hasti},
  booktitle={Proceedings of the 2025 Conference on Creativity and Cognition},
  pages={64--68},
  year={2025}
}

@inproceedings{lu2025research,
  title={Research on the Application of AI Painting Technology in Mobile Commerce Design},
  author={Lu, Yao and He, Sixuan and Zhong, Yuhan},
  booktitle={Proceedings of the 2025 International Conference on Generative Artificial Intelligence and Digital Media},
  pages={14--18},
  year={2025}
}

@inproceedings{xie2025embodied,
  title={Embodied Generative AI Art for Enhanced Human-Robot Interaction Through a Human-Centric LLM-Guided Robotic Arm Drawing System},
  author={Xie, Shengyuan and Sandoval, Eduardo Benitez and Shaik, Khaja Ahmed and Cruz, Francisco},
  booktitle={2025 20th ACM/IEEE International Conference on Human-Robot Interaction (HRI)},
  pages={1727--1730},
  year={2025},
  organization={IEEE}
}

@inproceedings{bran2023emerging,
  title={The emerging social status of generative AI: vocabularies of AI competence in public discourse},
  author={Bran, Emanuela and Rughini{\c{s}}, Cosima and Nadoleanu, Gheorghe and Flaherty, Michael G},
  booktitle={2023 24th International Conference on control systems and computer science (CSCS)},
  pages={391--398},
  year={2023},
  organization={IEEE}
}

@inproceedings{daniele2019ai+,
  title={AI+ art= human},
  author={Daniele, Antonio and Song, Yi-Zhe},
  booktitle={Proceedings of the 2019 AAAI/ACM Conference on AI, Ethics, and Society},
  pages={155--161},
  year={2019}
}

@inproceedings{brand2021design,
  title={A design inquiry into introspective AI: surfacing opportunities, issues, and paradoxes},
  author={Brand, Nico and Odom, William and Barnett, Samuel},
  booktitle={Proceedings of the 2021 ACM Designing Interactive Systems Conference},
  pages={1603--1618},
  year={2021}
}

@article{cetinic2022understanding,
  title={Understanding and creating art with AI: Review and outlook},
  author={Cetinic, Eva and She, James},
  journal={ACM transactions on multimedia computing, communications, and applications (TOMM)},
  volume={18},
  number={2},
  pages={1--22},
  year={2022},
  publisher={ACM New York, NY}
}

@article{amato2019ai,
  title={AI in the media and creative industries},
  author={Amato, Giuseppe and Behrmann, Malte and Bimbot, Fr{\'e}d{\'e}ric and Caramiaux, Baptiste and Falchi, Fabrizio and Garcia, Ander and Geurts, Joost and Gibert, Jaume and Gravier, Guillaume and Holken, Hadmut and others},
  journal={arXiv preprint arXiv:1905.04175},
  year={2019}
}

@book{zeilinger2021tactical,
  title={Tactical entanglements: AI art, creative agency, and the limits of intellectual property},
  author={Zeilinger, Martin},
  year={2021},
  publisher={meson press}
}

@article{kudless2023hierarchies,
  title={Hierarchies of bias in artificial intelligence architecture: Collective, computational, and cognitive},
  author={Kudless, Andrew},
  journal={International Journal of Architectural Computing},
  volume={21},
  number={2},
  pages={256--279},
  year={2023},
  publisher={SAGE Publications Sage UK: London, England}
}

@article{caramiaux2022explorers,
  title={" Explorers of Unknown Planets" Practices and Politics of Artificial Intelligence in Visual Arts},
  author={Caramiaux, Baptiste and Fdili Alaoui, Sarah},
  journal={Proceedings of the ACM on Human-Computer Interaction},
  volume={6},
  number={CSCW2},
  pages={1--24},
  year={2022},
  publisher={ACM New York, NY, USA}
}

@article{ch2019art,
  title={Art by computing machinery: Is machine art acceptable in the artworld?},
  author={Ch'ng, Eugene},
  journal={ACM Transactions on Multimedia Computing, Communications, and Applications (TOMM)},
  volume={15},
  number={2s},
  pages={1--17},
  year={2019},
  publisher={ACM New York, NY, USA}
}

@article{hertzmann2020computers,
  title={Computers do not make art, people do},
  author={Hertzmann, Aaron},
  journal={Communications of the ACM},
  volume={63},
  number={5},
  pages={45--48},
  year={2020},
  publisher={ACM New York, NY, USA}
}

@inproceedings{koch2020art,
  title={Where art meets technology: Integrating tangible and intelligent tools in creative processes},
  author={Koch, Janin and Pearson, Jennifer and Lucero, Andr{\'e}s and Sturdee, Miriam and Mackay, Wendy E and Lewis, Makayla and Robinson, Simon},
  booktitle={Extended Abstracts of the 2020 CHI Conference on Human Factors in Computing Systems},
  pages={1--7},
  year={2020}
}

@inproceedings{tang2024exploring,
  title={Exploring the impact of AI-generated image tools on professional and non-professional users in the art and design fields},
  author={Tang, Yuying and Zhang, Ningning and Ciancia, Mariana and Wang, Zhigang},
  booktitle={Companion Publication of the 2024 Conference on Computer-Supported Cooperative Work and Social Computing},
  pages={451--458},
  year={2024}
}

@inproceedings{lovato2024foregrounding,
  title={Foregrounding artist opinions: A survey study on transparency, ownership, and fairness in AI generative art},
  author={Lovato, Juniper and Zimmerman, Julia Witte and Smith, Isabelle and Dodds, Peter and Karson, Jennifer L},
  booktitle={Proceedings of the AAAI/ACM Conference on AI, Ethics, and Society},
  volume={7},
  pages={905--916},
  year={2024}
}

@inproceedings{yang2020re,
  title={Re-examining whether, why, and how human-AI interaction is uniquely difficult to design},
  author={Yang, Qian and Steinfeld, Aaron and Ros{\'e}, Carolyn and Zimmerman, John},
  booktitle={Proceedings of the 2020 chi conference on human factors in computing systems},
  pages={1--13},
  year={2020}
}

@inproceedings{park2024we,
  title={" We Are Visual Thinkers, Not Verbal Thinkers!": A Thematic Analysis of How Professional Designers Use Generative AI Image Generation Tools},
  author={Park, Hyerim and Eirich, Joscha and Luckow, Andre and Sedlmair, Michael},
  booktitle={Proceedings of the 13th Nordic Conference on Human-Computer Interaction},
  pages={1--14},
  year={2024}
}

@article{stork2024computer,
  title={Computer vision, ML, and AI in the study of fine art},
  author={Stork, David G},
  journal={Communications of the ACM},
  volume={67},
  number={5},
  pages={68--75},
  year={2024},
  publisher={ACM New York, NY, USA}
}

@inproceedings{bennett2024painting,
  title={Painting with Cameras and Drawing with Text: AI Use in Accessible Creativity},
  author={Bennett, Cynthia L and Shelby, Renee and Rostamzadeh, Negar and Kane, Shaun K},
  booktitle={Proceedings of the 26th International ACM SIGACCESS Conference on Computers and Accessibility},
  pages={1--19},
  year={2024}
}

@inproceedings{rezwana2025human,
  title={Human-Centered AI Communication in Co-Creativity: An Initial Framework and Insights},
  author={Rezwana, Jeba and Ford, Corey},
  booktitle={Proceedings of the 2025 Conference on Creativity and Cognition},
  pages={651--665},
  year={2025}
}

@inproceedings{davis2025co,
  title={The Co-Creative Design Framework for Hybrid Intelligence},
  author={Davis, Nicholas and Sherson, Jacob and Rafner, Janet},
  booktitle={Proceedings of the 2025 Conference on Creativity and Cognition},
  pages={560--572},
  year={2025}
}

@article{ma2024drawing,
  title={Drawing a satisfying picture: An exploratory study of human-AI interaction in AI Painting through breakdown--repair communication strategies},
  author={Ma, Xiaoyue and Huo, Yudi},
  journal={Information Processing \& Management},
  volume={61},
  number={4},
  pages={103755},
  year={2024},
  publisher={Elsevier}
}

@inproceedings{lima2025public,
  title={Public Opinions About Copyright for AI-Generated Art: The Role of Egocentricity, Competition, and Experience},
  author={Lima, Gabriel and Grgi{\'c}-Hla{\v{c}}a, Nina and Redmiles, Elissa M},
  booktitle={Proceedings of the 2025 CHI Conference on Human Factors in Computing Systems},
  pages={1--32},
  year={2025}
}

@article{kirova2023ethics,
  title={The ethics of artificial intelligence in the era of generative AI},
  author={Kirova, Vassilka D and Ku, Cyril S and Laracy, Joseph R and Marlowe, Thomas J},
  journal={Journal of Systemics, Cybernetics and Informatics},
  volume={21},
  number={4},
  pages={42--50},
  year={2023}
}

@inproceedings{ming2024labor,
  title={Labor, visibility, and technology: Weaving together academic insights and on-ground realities},
  author={Ming, Joy and Pei, Lucy and Varanasi, Rama Adithya and Kawakami, Anna and Verdezoto, Nervo and Cheon, EunJeong},
  booktitle={Companion Publication of the 2024 Conference on Computer-Supported Cooperative Work and Social Computing},
  pages={708--711},
  year={2024}
}

@inproceedings{knight2024impact,
  title={The impact of AI technology on the productivity of gig economy workers},
  author={Knight, Benjamin and Mitrofanov, Dmitry and Netessine, Serguei},
  booktitle={Proceedings of the 25th ACM Conference on Economics and Computation},
  pages={833--833},
  year={2024}
}

@inproceedings{munoz2023identity,
  title={Identity, Marginalization and Precarity in Platform-Mediated Freelancing},
  author={Munoz, Isabel},
  booktitle={Companion Proceedings of the 2023 ACM International Conference on Supporting Group Work},
  pages={69--71},
  year={2023}
}

@article{mako2022emerging,
  title={Emerging platform work in the context of the regulatory loophole (The Uber Fiasco in Hungary)},
  author={Mak{\'o}, Csaba and Ill{\'e}ssy, Mikl{\'o}s and Pap, J{\'o}zsef and Nosratabadi, Saeed},
  journal={Journal of Labor and Society},
  volume={26},
  number={4},
  pages={533--554},
  year={2022},
  publisher={Brill}
}

@article{vannoy2010social,
  title={The social influence model of technology adoption},
  author={Vannoy, Sandra A and Palvia, Prashant},
  journal={Communications of the ACM},
  volume={53},
  number={6},
  pages={149--153},
  year={2010},
  publisher={ACM New York, NY, USA}
}

@inproceedings{woodruff2024knowledge,
  title={How knowledge workers think generative ai will (not) transform their industries},
  author={Woodruff, Allison and Shelby, Renee and Kelley, Patrick Gage and Rousso-Schindler, Steven and Smith-Loud, Jamila and Wilcox, Lauren},
  booktitle={Proceedings of the 2024 CHI Conference on Human Factors in Computing Systems},
  pages={1--26},
  year={2024}
}

@article{zhang2024confrontation,
  title={" Confrontation or Acceptance": Understanding Novice Visual Artists' Perception towards AI-assisted Art Creation},
  author={Zhang, Shuning and Li, Shixuan},
  journal={arXiv preprint arXiv:2410.14925},
  year={2024}
}

@article{glikson2020human,
  title={Human trust in artificial intelligence: Review of empirical research},
  author={Glikson, Ella and Woolley, Anita Williams},
  journal={Academy of management annals},
  volume={14},
  number={2},
  pages={627--660},
  year={2020},
  publisher={Briarcliff Manor, NY}
}

@article{snow1987identity,
  title={Identity work among the homeless: The verbal construction and avowal of personal identities},
  author={Snow, David A and Anderson, Leon},
  journal={American journal of sociology},
  volume={92},
  number={6},
  pages={1336--1371},
  year={1987},
  publisher={University of Chicago Press}
}

@article{watson2008managing,
  title={Managing identity: Identity work, personal predicaments and structural circumstances},
  author={Watson, Tony J},
  journal={Organization},
  volume={15},
  number={1},
  pages={121--143},
  year={2008},
  publisher={Sage Publications Sage UK: London, England}
}

@article{stein2013towards,
  title={Towards an understanding of identity and technology in the workplace},
  author={Stein, Mari-Klara and Galliers, Robert D and Markus, M Lynne},
  journal={Journal of Information Technology},
  volume={28},
  number={3},
  pages={167--182},
  year={2013},
  publisher={SAGE Publications Sage UK: London, England}
}

@misc{discodiffusion,
  title        = {{Disco Diffusion}},
  author       = {{Alembics}},
  year         = {2021},
  howpublished = {\url{https://github.com/alembics/disco-diffusion}},
  note         = {Open-source CLIP-guided diffusion notebook for text-to-image generation. Accessed: 2026-05-15}
}

@misc{novelai,
  title        = {{NovelAI}},
  author       = {{Anlatan}},
  year         = {2022},
  howpublished = {\url{https://novelai.net/}},
  note         = {AI-assisted writing and anime-style image generation service. Accessed: 2026-05-15}
}

@article{zhang2020platform,
  title={When platform capitalism meets petty capitalism in China: Alibaba and an integrated approach to platformization},
  author={Zhang, Lin},
  journal={International Journal of Communication},
  volume={14},
  pages={21--21},
  year={2020}
}

@book{zhang2023labor,
  title={The labor of reinvention: Entrepreneurship in the new Chinese digital economy},
  author={Zhang, Lin},
  year={2023},
  publisher={Columbia University Press}
}

@article{scheen2019boredom,
  title={Boredom, Shanzhai, and digitisation in the time of creative China},
  author={Scheen, Lena and de Kloet, Jeroen and Chow, Yiu Fai},
  year={2019},
  publisher={Amsterdam University Press}
}

@article{yang2024rethinking,
  title={Rethinking the Copyrightability of Artificial Intelligence Generated Objects: Taking China's the AI Text-To-Picture Case as an Example},
  author={Yang, Zhezhe},
  journal={Journal of Economics and Law},
  volume={1},
  number={3},
  pages={20--28},
  year={2024}
}

@inproceedings{meng202652,
  title = {{52-Hz Whale Song: An Embodied VR Experience for Exploring Misunderstanding and Empathy}},
  author = {Meng, Yibo and Liu, Bingyi and Chen, Ruiqi and Chen, Xin and Guan, Yan},
  year = {2026},
  booktitle = {Proceedings of the Extended Abstracts of the 2026 CHI Conference on Human Factors in Computing Systems},
  pages = {1--5},
  publisher = {ACM},
  doi = {10.1145/3772363.3798690},
  url = {https://doi.org/10.1145/3772363.3798690}
}

@article{meng2026living,
  title = {{Living Inside the Black Box: Behavioral Probing and Adaptation in Mandatory Wearable Sensing}},
  author = {Meng, Yibo and Liu, Bingyi and Chen, Ruiqi and Ding, Xiaolan and Ma, Shuai},
  journal = {arXiv preprint arXiv:2607.09009},
  year = {2026},
  eprint = {2607.09009},
  archivePrefix = {arXiv},
  url = {https://arxiv.org/abs/2607.09009}
}

@article{meng2025tracing,
  title = {{Tracing Generative AI in Digital Art: A Longitudinal Study of Chinese Painters' Attitudes, Practices, and Identity Negotiation}},
  author = {Meng, Yibo and Chen, Ruiqi and Lu, Zhuoran and Ma, Shuai and Zang, Chengxi},
  journal = {arXiv preprint arXiv:2511.03117},
  year = {2025},
  eprint = {2511.03117},
  archivePrefix = {arXiv},
  url = {https://arxiv.org/abs/2511.03117}
}

@inproceedings{zeng2025parental,
  title = {{Parental Perceptions of Children’s d/Deaf Identity Shaping Technology Use: A Qualitative Study on Communication Technologies in Mixed-hearing Families}},
  author = {Zeng, Keyi and Lin, Jingyang and Chen, Ruiqi and LC, RAY and Hui, Pan and Tong, Xin},
  year = {2025},
  booktitle = {Proceedings of the Extended Abstracts of the CHI Conference on Human Factors in Computing Systems},
  pages = {1--10},
  publisher = {ACM},
  doi = {10.1145/3706599.3719753},
  url = {https://doi.org/10.1145/3706599.3719753}
}

@article{ma2026can,
  title = {{Can AI Agents Answer Your Data Questions? A Benchmark for Data Agents}},
  author = {Ma, Ruiying and Shankar, Shreya and Chen, Ruiqi and Lin, Yiming and Zeighami, Sepanta and Ghosh, Rajoshi and Gupta, Abhinav and Gupta, Anushrut and Gopal, Tanmai and Parameswaran, Aditya G.},
  journal = {arXiv preprint arXiv:2603.20576},
  year = {2026},
  eprint = {2603.20576},
  archivePrefix = {arXiv},
  url = {https://arxiv.org/abs/2603.20576}
}

@article{chen2026between,
  title = {{Between Knowledge and Care: A Mixed-Methods Evaluation of Generative AI for T2DM Self-Management from Patient and Physician Perspectives}},
  author = {Chen, Ruiqi and Meng, Yibo and Lu, Huidi and Ding, Xiaolan},
  journal = {arXiv preprint arXiv:2607.03720},
  year = {2026},
  eprint = {2607.03720},
  archivePrefix = {arXiv},
  url = {https://arxiv.org/abs/2607.03720}
}

@article{liu2025supporting,
  title = {{Supporting Our AI Overlords: Redesigning Data Systems to be Agent-First}},
  author = {Liu, Shu and Ponnapalli, Soujanya and Shankar, Shreya and Zeighami, Sepanta and Zhu, Alan and Agarwal, Shubham and Chen, Ruiqi and Suwito, Samion and Yuan, Shuo and Stoica, Ion and Zaharia, Matei and Cheung, Alvin and Crooks, Natacha and Gonzalez, Joseph E. and Parameswaran, Aditya G.},
  journal = {arXiv preprint arXiv:2509.00997},
  year = {2025},
  eprint = {2509.00997},
  archivePrefix = {arXiv},
  url = {https://arxiv.org/abs/2509.00997}
}

@inproceedings{zhang2025slideaudit,
  title = {{SlideAudit: A Dataset and Taxonomy for Automated Evaluation of Presentation Slides}},
  author = {Zhang, Zhuohao (Jerry) and Chen, Ruiqi and Zhong, Mingyuan and Wobbrock, Jacob O.},
  year = {2025},
  booktitle = {Proceedings of the 38th Annual ACM Symposium on User Interface Software and Technology},
  pages = {1--23},
  publisher = {ACM},
  doi = {10.1145/3746059.3747736},
  url = {https://doi.org/10.1145/3746059.3747736}
}

@inproceedings{luo2025s,
  title={" What's Happening"-A Human-centered Multimodal Interpreter Explaining the Actions of Autonomous Vehicles},
  author={Luo, Xuewen and Ding, Fan and Panda, Rishikesh and Chen, Ruiqi and Loo, Junnyong and Zhang, Shuyun},
  booktitle={Proceedings of the Winter Conference on Applications of Computer Vision},
  pages={1163--1170},
  year={2025}
}

@article{chen2026not,
  title = {{"Not Just Me and My To-Do List": Understanding Challenges of Task Management for Adults with ADHD and the Need for AI-Augmented Social Scaffolds}},
  author = {Chen, Jingruo and Meng, Yibo and Nie, Kexin},
  journal = {arXiv preprint arXiv:2603.17258},
  year = {2026},
  eprint = {2603.17258},
  archivePrefix = {arXiv},
  url = {https://arxiv.org/abs/2603.17258}
}

@article{meng2026balancing,
  title = {{Balancing Safety and Autonomy: Accessibility-Oriented Interventions in Generative AI for Cognitive Impairment}},
  author = {Meng, Yibo and Chen, Jingruo and Ye, Lyumanshan and Liu, Bingyi and Lu, Zhicong},
  journal = {arXiv preprint arXiv:2608.17175},
  year = {2026},
  eprint = {2608.17175},
  archivePrefix = {arXiv},
  url = {https://arxiv.org/abs/2608.17175}
}

@inproceedings{meng2026creating,
  title = {{"Creating the World with Order!': Designing Tangible Toolkit to Support Creative Expression and Wellbeing for Individuals with ASD}},
  author = {Meng, Yibo and Ye, Lyumanshan and Feng, Yifan and Fu, Rong and Liu, Bingyi and Li, Linghao and Gao, Nan},
  year = {2026},
  booktitle = {Proceedings of the 2026 Conference on Creativity and Cognition},
  pages = {666--683},
  publisher = {ACM},
  doi = {10.1145/3803784.3807531},
  url = {https://doi.org/10.1145/3803784.3807531}
}

@article{meng2026personalized,
  title = {{Personalized Intelligent Chatbot Based on AI-Generated Content Assists Memoir Writing for Older Adults With Cognitive Impairment: Mixed Methods Study}},
  author = {Meng, Yibo and Que, Yuan and Yan, Zhe and Liu, Bingyi and Wang, Zixin and Yang, Mandi and Lu, Huidi},
  year = {2026},
  journal = {JMIR Human Factors},
  volume = {13},
  pages = {e83428},
  publisher = {JMIR Publications Inc.},
  doi = {10.2196/83428},
  url = {https://doi.org/10.2196/83428}
}

\end{document}